# Novel analog switching circuit for van der Pauw measurements

Tal David[a]
Electrical Discharge and Plasma Laboratory, School of Physics and Astronomy, Tel Aviv University, Ramat Aviv, 69978, Israel

Itamar Molchadsky[b]
6 Modi'in Street, Tel Aviv, 62499, Israel

Avraham Somechi[c]
Computer-Electronic Student Laboratory, School of Physics and Astronomy, Tel Aviv University, Ramat Aviv, 69978, Israel

Ralph Rosenbaum[d]
School of Physics and Astronomy, Tel Aviv University, Raymond and Beverly Sackler Faculty of Exact Sciences, Ramat Aviv, 69978, Israel

## ABSTRACT

A simple electronic circuit is described using four common and very inexpensive analog multiplexer/demultiplexer chips. These analog switches are used to select eight different wiring configurations to a van der Pauw sample. Several interfacing schemes to a PC are suggested. The van der Pauw resistivity and Hall voltage expressions are also summarized.

## 1. INTRODUCTION

The van der Pauw technique is used to measure the resistivity and Hall voltages of *irregularly shaped* samples.[1,2] The technique requires eight different wiring configurations in order to extract the resistivity $\rho$ of the irregularly shaped sample.[1-4] Moreover, the sample should be uniform in thickness and contain no holes.
For the resistivity measurement, current and voltage leads have to be connected to four fixed contact terminals located on the edges of the sample using eight different wiring arrangements, as illustrated in Figs. 1 and 2. For the Hall voltage $V_{Hall}$ measurements, only two connecting arrangements are required as illustrated in Fig. 3; however, for each configuration, the magnetic field must be reversed to cancel unwanted resistive voltage contributions arising from the misalignment of the "diagonal Hall voltage terminals". In both cases, the rearrangement of the different connections becomes a tedious job if performed manually many times; it is time consuming to change the connections to the sample while still maintaining the same experimental conditions. This task is well suited for a PC. Four analog switches are required to accomplish this task. The switches are available commercially on plug-in cards for the PC, but these cards are expensive. We now describe an alternative solution using four identical and very inexpensive chips. In addition, the user will need either a data acquisition card (DAQ) or a GPIB (IEEE) 488 bus card installed in the mother board of the PC, along with two programmable digital voltmeters or two programmable ac lock-in amplifiers.

---

[a] Email: taldavid@bgu.ac.il
[b] Email: itamar@galooli.com
[c] Email: avrahams@post.tau.ac.il
[d] Email: Rachel-r@zahav.net.il

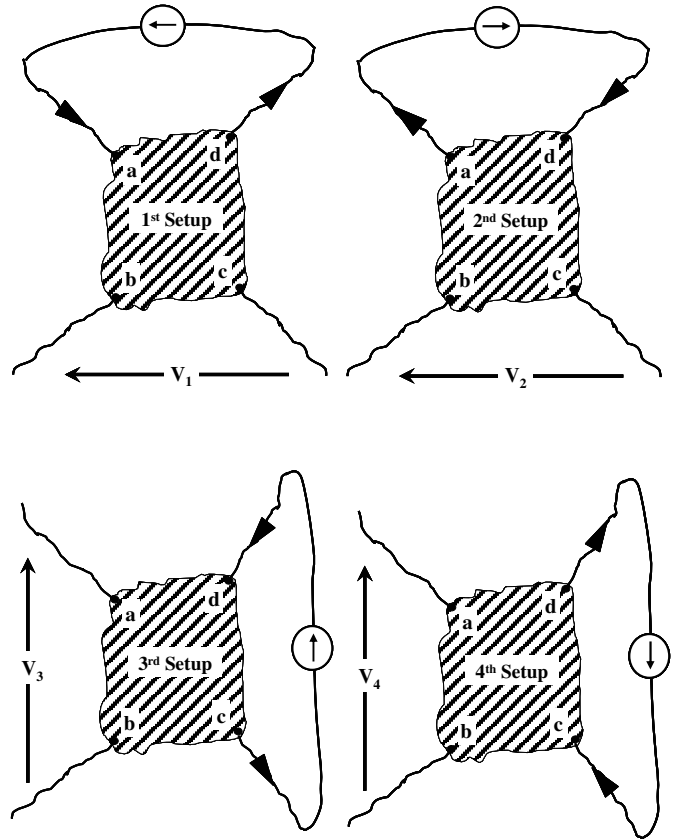

Fig 1. The four wiring configurations used for determining the value of the resistivity $\rho_A$.

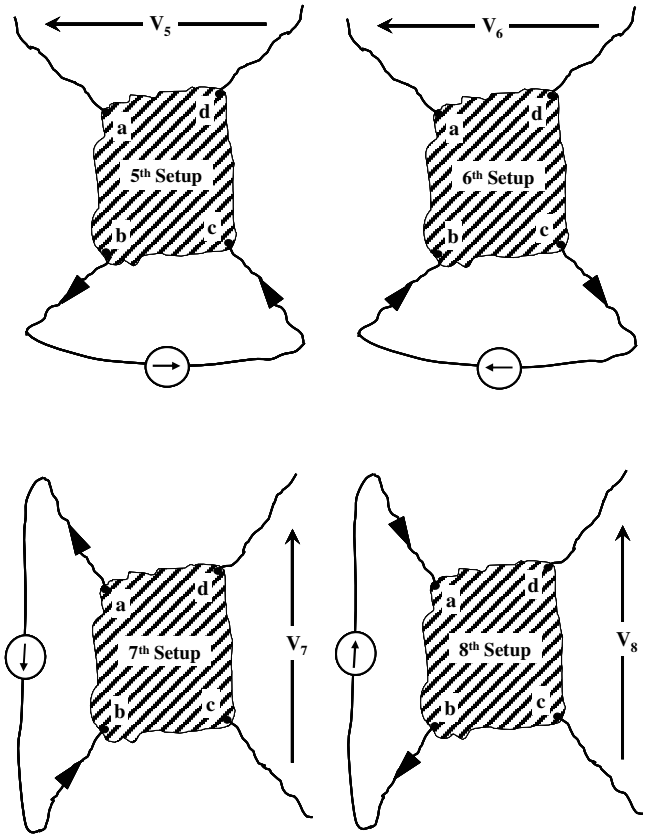

Fig 2. The four wiring configurations used for determining the value of the resistivity $\rho_B$.

## 2. RESISTIVITY AND HALL VOLTAGE EXPRESSIONS

The van der Pauw equations have been nicely summarized in the book, <u>Low Level Measurements</u>, by Keithley Instruments, Inc.[3], which we now summarize. Also refer to the publication written by the National Institute of Standards and Technology.[4]

As shown in Figs. 1 and 2, a total of eight voltage measurements are required. Two resistivities, $\rho_A$ and $\rho_B$, are then evaluated as follows:

$$\rho_A = 1.133\, f_A \cdot t_s (V_2 - V_1 + V_4 - V_3)/I$$

and

$$\rho_B = 1.133\, f_B \cdot t_s (V_6 - V_5 + V_8 - V_7)/I \quad , \tag{1}$$

where $\rho_A$ and $\rho_B$ are resistivities in $\Omega \cdot cm$, $t_s$ is the sample thickness in cm, $V_1$ through $V_8$ are the measured voltages defined in Figs. 1 and 2, $I$ is the current in Amp, and $f_A$ and $f_B$ are geometrical factors based on sample symmetry and are related to the two resistance ratios $Q_A$ and $Q_B$ as outlined below ($f_A = f_B = 1$ for perfect symmetry). Note that van der Pauw suggested that the prefactor should be 2.266 as given in his Eq. (11)[1] and not 1.133 as given above by the Keithley Instrument group.[3]

$Q_A$ and $Q_B$ can be calculated using the measured voltages as follows:

$$Q_A = (V_2 - V_1)/(V_4 - V_3)$$

and

$$Q_B = (V_6 - V_5)/(V_8 - V_7) \ . \tag{2}$$

Also the $Q$'s and $f$'s are related as follows:

$$(Q - 1)/(Q + 1) = (f/0.693) \operatorname{arccosh}[0.5\exp(0.693/f)] \ . \tag{3}$$

There appears to be a typographical mistake in van der Pauw's article[1] as the prefactor appearing on the right side of our Eq. (3), the 1/0.693 term, is missing in his Eq. (3).[1] However, both van der Pauw and Keithley give identical plots of $f$ versus $Q$. A condensed plot of $f$ versus $Q$ is given in Fig. 4.

Once $\rho_A$ and $\rho_B$ are known, the average resistivity $\rho_{ave}$ can be calculated as follows:

$$\rho_{ave} = (\rho_A + \rho_B)/2 \ . \tag{4}$$

Note that if $\rho_A$ and $\rho_B$ are not within 10 % of one another, the sample is not sufficiently uniform to accurately determine its resistivity. Refer to Ref. 4 for details.

The Hall voltage $V_{Hall}$ and the corresponding carrier concentration $n$ or $p$ (carriers/m$^3$) are measured by applying a magnetic field perpendicular to the sample.[4] By referring to Fig. 3 for the definitions of the measured voltages $V_A$ through $V_D$, the Hall voltage is:

$$V_{Hall} = (V_A - V_B + V_C - V_D)/4 \ , \tag{5}$$

and the carrier concentration is:

$$n \text{ or } p = IB/(e \cdot t_s \cdot V_{Hall}) \ . \tag{6}$$

Note that the sign of the carrier (electrons or holes) can be determine by a careful dc measurement of both the sign and magnitude of the Hall voltage.

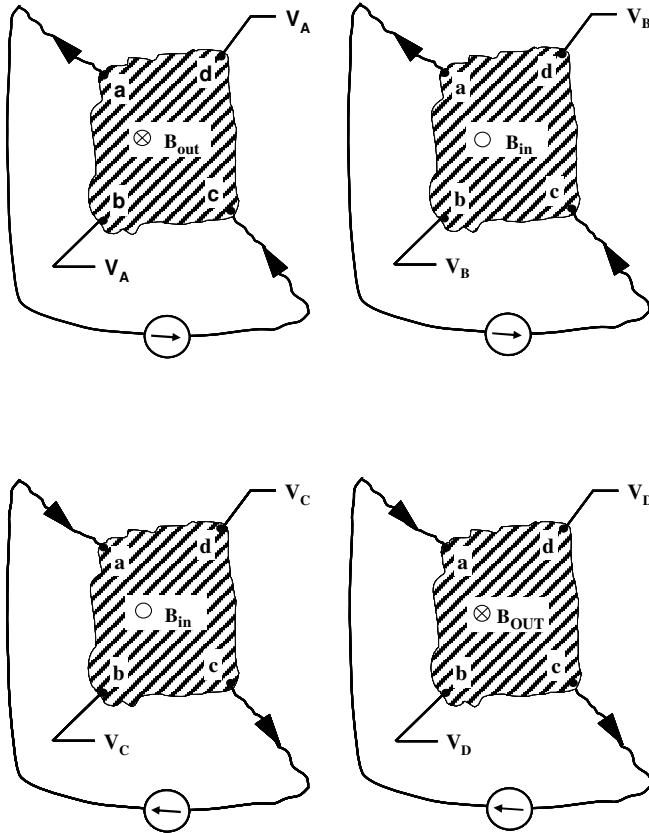

Fig 3. The *two* wiring configurations used for determining the value of the Hall voltage $V_{Hall}$. Notice the need to reverse the perpendicular magnetic field direction in each configuration.

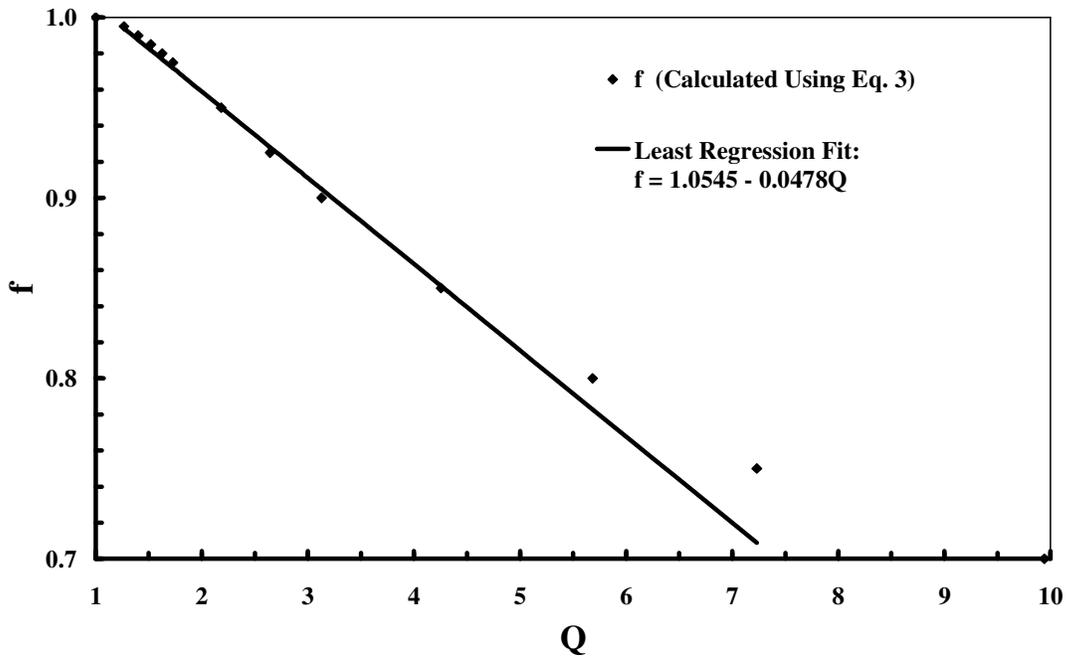

Fig 4. Plot of *f* (defined by Eq. 3) versus *Q* (defined by Eq. 2).

## 3. THE ANALOG SWITCHES

The MC74HC4502A-N "analog multiplexer/demultiplexer" chip from ON Semiconductors is well suited to this application.[5] Each chip has two bidirectional switches, namely this chip is a double-pole, 4-position device with a common off. Unfortunately, each switch section cannot be programmed individually as shown in its data sheet given in Fig. 5[5]; thus, we can use only one of the two switches that are available on each chip. We need four different chips: one to inject the current, the second to extract the current, and the third and fourth units to select the corresponding differential voltage.

Each chip has two control lines to select one of the four input/output legs, X0, X1, X2 or X3 (namely, one of the four terminals on the sample - a, b, c, or d) and also an enable line; refer to Fig. 5 for wiring details.[5] We recommend that each chip be disenable while selecting one of its legs (or one terminal of the sample). Thus there are three control lines that program the configuration each switch to the sample. And because there are four different chips being used, there are a total of twelve control lines. These twelve control lines can be programmed using a Digital Input/Output feature of a Data Acquisition Card (DAQ). But recall that many DAQ cards have only 8 lines, too few for this application. Thus, it is important that the DAQ card have at least 12 digital input/output lines. Alternatively, one can purchase the inexpensive Digital Input/Output card, the PIC-DI024, from Measurement Computing Corporation which has 24 lines.[6]

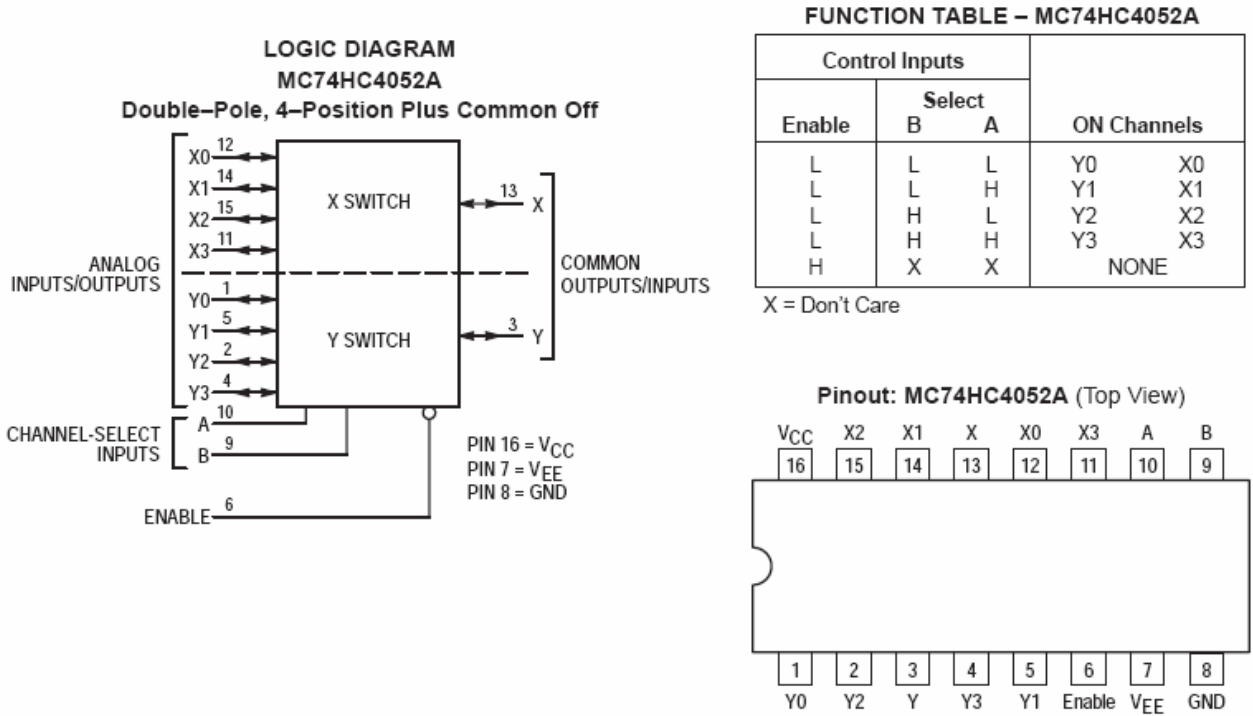

Fig 5. Logic table and diagram of the control legs of the analog multiplexer/demultiplexer chip, the MC74HC4052A, taken from the data sheets of ON Semiconductor.

The MC74HC4502A chip has these following useful properties:[5]
   a) Power supply voltages: $V_{cc}$ = + 5 V and $V_{EE}$ = -5 V;
   b) TTL logic for the two "Channel-Select" lines and for the chip Enable line;
   c) Analog voltage extremes between +5 to -5 V;
   d) Cross talk between input lines: about -50 dB's or 1 to 100,000 ratio;
   e) Leakage current of an input line to ground: about 0.1 µA;
   f) Useful frequency range up to 80 MHz;
   g) Cost: a fraction of one US$.

The MC74HC4502A chip has these disadvantages compared to a mechanical switch:[5]
   1) Typical resistance of 30 Ω to 35 Ω compared to a small fraction of an Ω;
   2) Maximum current through switch is limited to about ±25 mAmps compared to Amps or a fraction of an Amp.

For samples with sheet resistances of 1 kΩ/ to 10 kΩ/ that we have measured, these two limitations are unimportant.

A suggested circuit diagram is illustrated in Fig. 6. The four switches are not shown in this diagram in order to simplify the drawing. We assume that the current and voltage wires are already connected correctly to the sample using the four analog switches. There is a standard reference resistor $R_{current}$ also placed in series with the circuit; its use is optional. This resistor measures the current $I$, if its resistance $R_{current}$ is known; thus, the voltage drop across this resistor needs to be measured. In addition, it can be used to limit the total current flowing through the circuit if small currents on the order of µAmps are required.

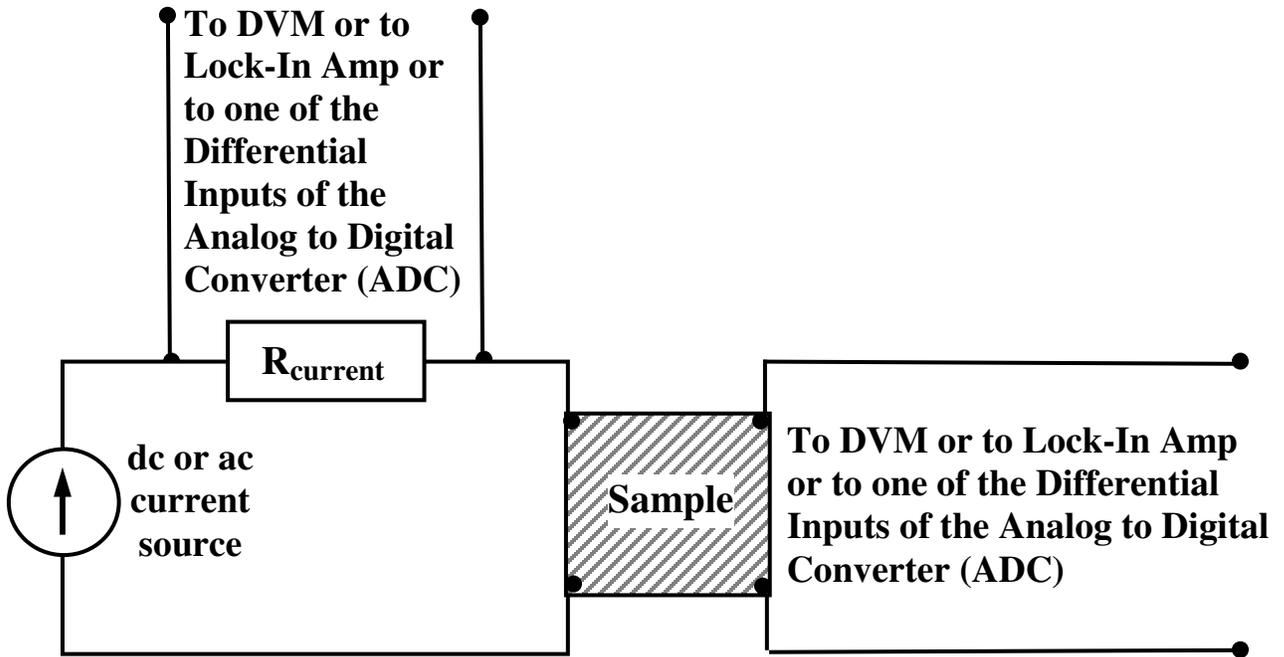

Fig 6. General circuit diagram for interfacing the van der Pauw sample to the PC.

There are several routes to interface the measured voltages to the PC, depending upon the hardware and software available in your laboratory. If you have an appropriate data acquisition card (DAQ), you can use the differential voltage mode of its analog-digital converter (ADC) and its multi-input selection feature to measure either the van der Pauw voltage or the voltage drop across the standard resistor. But if you measure Hall voltages, there could be a problem. Recall that a 12 bit converter can measure at best only about 1 mV at its best resolution. This resolution is insufficient if you are measuring Hall voltages that are on the order of µV's, which are typical for a metal having a large carrier number density. Alternatively, you can replace the ADC section with two programmable *microvolt* DVM's and can read their outputs using either the GPIB bus or the RS232 bus. Another possibility is to use two lock-in amplifiers and to detect the minutely small ac Hall voltages resulting from an ac current flowing through the sample and standard resistor. Again each lock-in amplifier can be read using either the GPIB or RS232 buses. However, the lock-in amplifier cannot uniquely determine the sign of the carrier owing to the arbitrary $180^0$ phase setting, used in detecting the maximum *magnitude* of the resistive component of the Hall voltage.

Thus, there are many different circuit arrangements that can be used together with these neat analog switches to interface the van der Pauw sample to the PC. In the references, we list several commercial companies who sell programmable instruments, DAQ cards and Digital Input/Output cards.[3,6-8] Some useful software programs are National Instruments's LabVIEW or Agilent's VEE programs for controlling the GPIB bus and the cards installed on the mother board of the PC.[7,9,3] Our listing is far from complete. We do not endorse any particular company; we have had good experiences using the products from all these manufacturers.

## ACKNOWLEDGEMENTS

We greatly thank Prof. Samuel Goldsmith of the School of Physics and Prof. Ray Boxman of the Faculty of Engineering at TAU for their guidance and supervision. We are obliged to Mrs. Rachel Rosenbaum for her graphical work. We acknowledge Dr. Benny D'wir of the Ecole Polytechnique Federale de Lausanne for his Hall voltage expression.